# Decoherence of near-surface nitrogen-vacancy centers due to electric field noise

M. Kim[1], H. J. Mamin[1], M. H. Sherwood[1], K. Ohno[2,+], D. D. Awschalom[2] and D. Rugar[1,*]

[1]IBM Research Division, Almaden Research Center, San Jose, CA 95120, USA
[2]Institute for Molecular Engineering, University of Chicago, IL 60637, USA
[+]Current address: Applied Materials, Inc., 3340 Scott Blvd., Santa Clara, CA 95054, USA

*For correspondence: rugar@us.ibm.com

**Abstract**

We show that electric field noise from surface charge fluctuations can be a significant source of spin decoherence for near-surface nitrogen-vacancy (NV) centers in diamond. This conclusion is based on the increase in spin coherence observed when the diamond surface is covered with high-dielectric-constant liquids, such as glycerol. Double resonance experiments show that improved coherence occurs even though the coupling to nearby electron spins is unchanged when the liquid is applied. Multipulse spin echo experiments reveal the effect of glycerol on the spectrum of NV frequency noise.

The negatively charged nitrogen-vacancy (NV) center in diamond is attracting great interest as an atomic-size quantum sensor that is operable at room temperature and has a convenient readout via optical fluorescence. NV centers are finding wide ranging applications due to their responsiveness to local magnetic [1,2], electric [3,4], strain [5,6] and temperature fields [7,8]. In most cases, the sensitivity of the NV center is critically dependent on the long quantum coherence time of its spin state, which in bulk diamond can be greater than 1 ms at room temperature [9].

In many nanoscale sensing applications the NV center must be located as close to the surface as possible in order to maximize the detected signal [10–14]. Unfortunately, significant impairment of the spin coherence has been found for NV centers located within a few nanometers of the diamond surface [15–18]. In the NV-diamond research community, this near-surface decoherence is commonly attributed to magnetic noise emanating from unpaired electron spins in surface dangling bonds [15–19].

In this paper we present evidence that near-surface NV decoherence is not solely due to magnetic noise, but instead can be dominated by electric field noise from surface charge fluctuations. This finding is based on the improvement of coherence seen when high-dielectric-constant liquids are applied to the diamond surface. For example, when the diamond is immersed in glycerol, we have found that Hahn echo $T_2$ times can increase by more than a factor of four. To rule out the influence of magnetic noise due to surface spins, we directly probed the surface electron spin density with a double resonance experiment and found no significant change upon application of the glycerol. With simple electrostatic calculations, combined with the known NV spin Hamiltonian, we show that decoherence due to charge fluctuations is physically reasonable. Finally, we use the results from multipulse dynamic decoupling experiments to estimate the spectral density of the NV frequency noise.

Our experiments were performed using an electronic grade (100)-oriented diamond substrate that was capped with a 50 nm thick layer of isotopically pure carbon-12 diamond. Near-surface NV centers were created by [15]N ion implantation at 2.5 keV, followed by annealing in vacuum at 850°C, acid cleaning and heating to 425°C in a pure oxygen atmosphere [20]. This process results in NV centers located at depths roughly 5 nm below the surface. Individual NV centers were detected by confocal fluorescence microscopy with photon counting electronics. The custom built microscope had an inverted geometry that incorporated a small windowed cell which allowed liquid to be applied to the top surface of the diamond. See Supplemental Material for further details on sample preparation and apparatus [21].

Optically detected spin echo experiments were performed with applied magnetic field in the range of 20 – 40 mT directed along the [111] symmetry axis of the NV center (Fig. 1(a)). Measurements were made both before and after applying various liquids to the diamond surface. Four liquids were tested: conventional and fully deuterated glycerols (dielectric constant $\kappa_G = 42$), propylene carbonate ($\kappa_{PC} = 64$) and microscope immersion oil ($\kappa_{oil} = 2.3$). We note that glycerol and propylene carbonate have quite different chemical characteristics. Glycerol is an alcohol whose hydroxyl groups can donate protons to the environment, possibly leading to some passivation of surface dangling bonds. In contrast, propylene carbonate is known to be an aprotic solvent, meaning that the hydrogen atoms of the molecule are tightly bound.

As shown in Fig. 1(b), a dramatic 4.6× increase in $T_2$ was found when deuterated glycerol was placed on the diamond surface. Such a large increase indicates that the noise responsible for NV decoherence was substantially suppressed when the glycerol was added. After the glycerol was removed and the diamond recleaned, propylene carbonate was applied, again resulting in a significant 2.4× increase in $T_2$ time (Fig. 1(c)). In contrast, when the same NV center was studied with immersion oil, only a small 1.4× increase in $T_2$ was observed (Fig. 1(d)). Similar comparisons were performed with six other NV centers, with results summarized in Fig. 1(e). Substantial improvements in NV coherence were found when any of the three high-$\kappa$ liquids were applied to the diamond surface, with $T_2$ ratios ($T_{2,\text{Liquid}} / T_{2,\text{Air}}$) ranging from 1.7 to 4.6. In contrast, application of the lower $\kappa$ immersion oil showed little or no coherence improvement, with $T_2$ ratios ranging from 0.8 to 1.4.

To test whether the passivation of surface electron spins ("dark spins") is a possible mechanism of coherence improvement, we performed a double electron-electron resonance (DEER) experiment (Fig. 2(a)) [22,23]. We measured the NV spin echo while applying an additional microwave pulse half-way through the spin echo sequence. This pulse inverts the dark spins when its frequency is resonant with the dark spin precession frequency. The inversion of dark spins that are in close proximity to the NV causes a change the local magnetic field at the NV center and results in a dip in the echo response.

As can be seen in Fig. 2(b), a clear dip in the spin echo signal occurs when the frequency of the microwave pulse matches the resonance frequency of the $g \approx 2$ dark spins (1.09 GHz in a 39 mT field), indicating that unpaired electron spins are indeed present in the neighborhood of the NV center. When the experiment was repeated after the addition of deuterated glycerol, the $T_2$ time increased by a factor of 2.4, but the DEER signal was essentially unchanged. Since no significant change is seen, it appears that the surface electron spin density is largely unaffected by the addition of the glycerol, and thus not the key factor in the observed $T_2$ improvement.

Given the DEER results above, we conclude that the improvement of coherence time with glycerol and propylene carbonate is most likely related to the high dielectric constants of these liquids, suggesting that much of the near-surface NV decoherence is the result of electric field noise due to fluctuating surface charges. A simple electrostatic calculation illustrates the action of the high dielectric constant liquid. Consider a point charge $q$ on the surface of the diamond. The resulting electric field at the NV center depends on the dielectric constants of both the diamond and the external medium according to [24]

$$\mathbf{E} = \frac{1}{4\pi\varepsilon_0} \frac{2}{\kappa_d + \kappa_{ext}} \frac{q}{r^2} \hat{\mathbf{r}}, \tag{1}$$

where $r$ is the distance between the surface charge and the NV center, $\hat{\mathbf{r}}$ is the unit vector in the direction of the NV center, $\kappa_d = 5.7$ is the dielectric constant of diamond, $\kappa_{ext}$ is the dielectric constant external to the diamond and $\varepsilon_0$ is the permittivity of free space. Compared to a diamond in air, the reduction of electric field when an external medium (the liquid) is applied is given by $E/E_{air} = (\kappa_d + 1)/(\kappa_d + \kappa_{ext})$. For the case of glycerol with $\kappa_{ext} = 42$, the electric field is thus reduced by a factor of 7. For a single electronic charge on the diamond surface, the electric field for a NV located 5 nm below the charge is $1.7 \times 10^7$ V/m when the diamond is in air, and reduced to $2.4 \times 10^6$ V/m with glycerol on the surface.

To show that fluctuating electric fields on the order of $10^7$ V/m are sufficient to cause significant decoherence, we start with the NV spin Hamiltonian [3,25]

$$H = \left(hD + d_{\parallel}E_z\right)\left[S_z^2 - 2/3\right] + \mu_B g_{NV} \mathbf{S} \cdot \mathbf{B} - d_{\perp}\left[E_x\left(S_xS_y + S_yS_x\right) + E_y\left(S_x^2 - S_y^2\right)\right] \tag{2}$$

where $h$ is Planck's constant, $\mathbf{S}$ is the $S=1$ electron spin operator, $\mathbf{B}$ is the applied magnetic field, $\mathbf{E}$ is the electric field at the NV center, $D = 2.87$ GHz is the zero field splitting, $\mu_B$ is the Bohr magneton and $g_{NV} \approx 2$ is the electron spin g-factor. The electric field acts on the NV center via the coupling parameters $d_\parallel / h = 3.5$ mHz m V$^{-1}$ and $d_\perp / h = 170$ mHz m V$^{-1}$. To find the effect of electric field on the NV spin precession frequency, we assume the applied magnetic field is aligned with the NV symmetry axis (z axis). We can then solve for the energy eigenvalues associated with the three magnetic sublevels $m_s = +1, 0$ or $-1$ and find the precession frequencies for superpositions between the $m_s = 0$ and the $\pm 1$ states. The resulting change in precession frequency due to an electric field is [3]

$$\Delta \omega_\pm / 2\pi = (d_\parallel / h) E_z \pm \frac{1}{2} \frac{(d_\perp / h)^2 E_\perp^2}{(\gamma / 2\pi) B_z} \qquad (3)$$

where $E_\perp^2 = E_x^2 + E_y^2$, $\gamma / 2\pi = g_{NV} \mu_B / h = 28$ GHz/T, and we have assumed that $(d_\perp E_\perp / g_{NV} \mu_B B_z)^2 \ll 1$.

With equation (3) we can now determine the frequency shift due to a single elementary charge located directly above a 5 nm deep NV center. Assuming a magnetic field of 20 mT and a (100)-oriented diamond substrate, where the NV z-axis is tilted by 54.7° with respect to the surface normal, the $E_z$ term contributes a 35 kHz shift. The $E_\perp$ term contributes an additional ±5 kHz, for a total frequency shift of up to 40 kHz. Frequency fluctuations of this magnitude would be sufficient to give a dephasing time $T_2^* \sim 1/|\Delta \omega|$ in the range of microseconds to tens of microseconds, depending on the spectrum of the fluctuations. While the effect of $E_\perp$ is fairly modest in this example, it becomes relatively more important the larger the electric field (i.e., when more charges are present and for shallower NVs) since it contributes quadratically in (3).

To better understand the frequency spectrum of the fluctuations that cause the near-surface decoherence, we performed multipulse dynamic decoupling experiments (Fig. 3(a)) [16,18,26,27]. NVs were studied both before and after the application of deuterated glycerol using XY8-N pulse sequences [11,28], where N is the number of π pulses in the sequence (N = 1, 32, 96 and 256). Figures 3(b) and 3(c) show spin coherence data as a function of total evolution time based on measured spin echo amplitudes. The curves were found to be well fit by stretched exponentials of the form $\exp\left[-(t/T_2)^n\right]$. As expected, the $T_2$ times increased with the number of π pulses (Fig. 3(d)), and exhibited the power law dependence $T_2 \propto N^k$, with $k$ = 0.52 for the air case and 0.41 with glycerol.

The coherence data in Fig. 3 can be used to estimate the spectrum of NV frequency fluctuations by taking into account the filter functions associated with the decoupling sequences [29]. Using a spectral decomposition procedure similar to that described in Refs. [18] and [27], we extract an estimate for the spectral density of the NV precession frequency, $S_\omega(\omega)$ [30]. In air, the spectrum roughly fits a $1/\omega$ dependence between 10 kHz and 1 MHz (Fig. 4). The addition of glycerol substantially reduces the spectral density for frequencies between 10 kHz and 100 kHz, where a dependence of $\sim 1/\omega^{0.8}$ is seen. It is this reduction of spectral density that is most responsible for the observed increase in $T_2$ times. Above 100 kHz, the glycerol spectrum flattens out, and beyond 600 kHz, the spectral density with glycerol is approximately equal to the spectral density without glycerol.

The ineffectiveness of glycerol to cancel electric field noise above 600 kHz is somewhat surprising given that the dielectric relaxation frequency for bulk liquid glycerol has been measured to be greater than 100 MHz [31]. One possibility is that the dielectric relaxation frequency is much reduced at the surface of the diamond. For example, experiments probing nanoscale layers of glycerol on surfaces have found evidence that a nanometer-thick layer of reduced mobility can form at the solid-liquid interface [32–34]. This semi-solid layer could impede the rotation of glycerol molecules and thereby reduce the effective dielectric constant at high frequencies. A second possibility is that thermal agitation of the glycerol molecules adds broadband electric field noise and thereby sets a floor to the spectral density that becomes the dominant noise source at higher frequencies. A straightforward calculation shows that randomly rotating electric dipoles from glycerol molecules will create a substantial fluctuating electric field of approximately $10^7$ V/m-rms at a depth of 5 nm.

It is tempting to use our estimate of $S_\omega(\omega)$ to find the electric field spectral density. Unfortunately, with our current dataset, the frequency mixing behavior of the $E_\perp^2$ nonlinearity in (3) makes it impossible to rigorously determine the electric field spectral densities without making some significant assumptions about the noise spectrum in frequency regions where we have no direct experimental information. For example, a substantial DC electric field from static surface charge would not be directly evident in our measurements, but would act to enhance the relative contribution of fluctuating fields via the $E_\perp^2$ nonlinearity.

If we take a naïve approach and consider only the $E_z$ contribution in (3), then the analysis is straightforward and we can write $S_\omega = 4\pi^2 (d_\parallel / h)^2 S_{E_z}$. To find the electric field spectral density $S_{E_z}$ we take $S_\omega$ in air from Fig. 4, which is approximately $S_\omega = 1.4 \times 10^{10} \text{ s}^{-2} / |\omega|$, and obtain $S_{E_z} = 2.9 \times 10^{13} (\text{V/m})^2 / |\omega|$. Integrating this over the measured range of 10 kHz to 1 MHz, we find $\langle E_z^2 \rangle^{1/2} = 6.5 \times 10^6$ V/m, which is less than the equivalent of one electronic charge at 5 nm distance. This value should be viewed as a very conservative lower bound to the total fluctuating field since we are considering only one vector component of field and over a very limited frequency range.

In closing, we note that an alternative approach for distinguishing between electric and magnetic field noise in NV decoherence is to compare conventional spin echo results with "double-quantum" spin echoes, which utilize the superposition between the $m_s$ = -1 and +1 sublevels [35–38]. We explore this avenue in the Supplementary Material [21] and show results that support our conclusion that electric field noise can be a significant contributor to decoherence for near-surface NV centers.

The authors thank B. Myers, A. Jayich, M. Salmeron and J. Hodges for helpful discussions. This work was supported by the DARPA QuASAR program and the US Air Force Office of Scientific Research.

# References

[1] G. Balasubramanian, I. Y. Chan, R. Kolesov, M. Al-Hmoud, J. Tisler, C. Shin, C. Kim, A. Wojcik, P. R. Hemmer, A. Krueger, T. Hanke, A. Leitenstorfer, R. Bratschitsch, F. Jelezko, and J. Wrachtrup, Nature **455**, 648 (2008).

[2] J. R. Maze, P. L. Stanwix, J. S. Hodges, S. Hong, J. M. Taylor, P. Cappellaro, L. Jiang, M. V. G. Dutt, E. Togan, A. S. Zibrov, A. Yacoby, R. L. Walsworth, and M. D. Lukin, Nature **455**, 644 (2008).

[3] F. Dolde, H. Fedder, M. W. Doherty, T. Nöbauer, F. Rempp, G. Balasubramanian, T. Wolf, F. Reinhard, L. C. L. Hollenberg, F. Jelezko, and J. Wrachtrup, Nat. Phys. **7**, 459 (2011).

[4] F. Dolde, M. W. Doherty, J. Michl, I. Jakobi, B. Naydenov, S. Pezzagna, J. Meijer, P. Neumann, F. Jelezko, N. B. Manson, and J. Wrachtrup, Phys. Rev. Lett. **112**, 097603 (2014).

[5] P. Ovartchaiyapong, K. W. Lee, B. A. Myers, and A. C. B. Jayich, Nat. Commun. **5**, 4429 (2014).

[6] E. R. MacQuarrie, T. A. Gosavi, N. R. Jungwirth, S. A. Bhave, and G. D. Fuchs, Phys. Rev. Lett. **111**, 227602 (2013).

[7] D. M. Toyli, C. F. de las Casas, D. J. Christle, V. V. Dobrovitski, and D. D. Awschalom, Proc. Natl. Acad. Sci. USA **110**, 8417 (2013).

[8] G. Kucsko, P. C. Maurer, N. Y. Yao, M. Kubo, H. J. Noh, P. K. Lo, H. Park, and M. D. Lukin, Nature **500**, 54 (2013).

[9] G. Balasubramanian, P. Neumann, D. Twitchen, M. Markham, R. Kolesov, N. Mizuochi, J. Isoya, J. Achard, J. Beck, J. Tissler, V. Jacques, P. R. Hemmer, F. Jelezko, and J. Wrachtrup, Nat. Mater. **8**, 383 (2009).

[10] H. J. Mamin, M. Kim, M. H. Sherwood, C. T. Rettner, K. Ohno, D. D. Awschalom, and D. Rugar, Science **339**, 557 (2013).

[11] T. Staudacher, F. Shi, S. Pezzagna, J. Meijer, J. Du, C. A. Meriles, F. Reinhard, and J. Wrachtrup, Science **339**, 561 (2013).

[12] M. S. Grinolds, S. Hong, P. Maletinsky, L. Luan, M. D. Lukin, R. L. Walsworth, and A. Yacoby, Nat. Phys. **9**, 215 (2013).

[13] D. Rugar, H. J. Mamin, M. H. Sherwood, M. Kim, C. T. Rettner, K. Ohno, and D. D. Awschalom, Nat. Nanotech. **10**, 120 (2015).

[14] T. Häberle, D. Schmid-Lorch, F. Reinhard, and J. Wrachtrup, Nat. Nanotech. **10**, 125 (2015).

[15] T. Rosskopf, A. Dussaux, K. Ohashi, M. Loretz, R. Schirhagl, H. Watanabe, S. Shikata, K. M. Itoh, and C. L. Degen, Phys. Rev. Lett. **112**, 147602 (2014).

[16] B. A. Myers, A. Das, M. C. Dartiailh, K. Ohno, D. D. Awschalom, and A. C. Bleszynski Jayich, Phys. Rev. Lett. **113**, 027602 (2014).

[17] B. K. Ofori-Okai, S. Pezzagna, K. Chang, M. Loretz, R. Schirhagl, Y. Tao, B. A. Moores, K. Groot-Berning, J. Meijer, and C. L. Degen, Phys. Rev. B **86**, 081406(R) (2012).

[18] Y. Romach, C. Müller, T. Unden, L. J. Rogers, T. Isoda, K. M. Itoh, M. Markham, A. Stacey, J. Meijer, S. Pezzagna, B. Naydenov, L. P. McGuinness, N. Bar-Gill, and F. Jelezko, Phys. Rev. Lett. **114**, 017601 (2015).

[19] L. Luan, M. S. Grinolds, S. Hong, P. Maletinsky, R. L. Walsworth, and A. Yacoby, Sci. Rep. **5**, 8119 (2015).

[20] M. Kim, H. J. Mamin, M. H. Sherwood, C. T. Rettner, J. Frommer, and D. Rugar, Appl. Phys. Lett. **105**, 042406 (2014).

[21] See Supplemental Material at http:/link.aps.org/supplemental/xxxxx for details on methods, apparatus and supporting experiments.

[22] B. Grotz, J. Beck, P. Neumann, B. Naydenov, R. Reuter, F. Reinhard, F. Jelezko, J. Wrachtrup, D. Schweinfurth, B. Sarkar, and P. Hemmer, New J. Phys. **13**, 055004 (2011).

[23] H. J. Mamin, M. H. Sherwood, and D. Rugar, Phys. Rev. B **86**, 195422 (2012).

[24] J. D. Jackson, *Classical Electrodynamics* (Wiley, New York, 1975), pp. 147-9.

[25] E. Van Oort and M. Glasbeek, Chem. Phys. Lett. **168**, 529 (1990).

[26] G. de Lange, Z. H. Wang, D. Ristè, V. V Dobrovitski, and R. Hanson, Science **330**, 60 (2010).

[27] N. Bar-Gill, L. M. Pham, C. Belthangady, D. Le Sage, P. Cappellaro, J. R. Maze, M. D. Lukin, A. Yacoby, and R. Walsworth, Nat. Commun. **3**, 858 (2012).

[28] T. Gullion, D. B. Baker, and M. S. Conradi, J. Magn. Res. **89**, 479 (1990).

[29] Ł. Cywinski, R. M. Lutchyn, C. P. Nave, and S. Das Sarma, Phys. Rev. B **77**, 174509 (2008).

[30] We assume a double-sided convention for the spectral density with the normalization $(1/2\pi)\int_{-\infty}^{\infty} S_\omega(\omega) d\omega = \langle (\Delta\omega)^2 \rangle$, where $\langle (\Delta\omega)^2 \rangle$ is the mean square of the frequency fluctuations.

[31] K. L. Ngai, P. Lunkenheimer, C. León, U. Schneider, R. Brand, and A. Loidl, J. Chem. Phys. **115**, 1405 (2001).

[32] S. Capponi, S. Napolitano, N. R. Behrnd, G. Couderc, J. Hulliger, and M. Wübbenhorst, J. Phys. Chem. C **114**, 16696 (2010).

[33] L. Xu and M. Salmeron, J. Phys. Chem. B **102**, 7210 (1998).

[34] With both glycerol and propylene carbonate we have frequently seen the $T_2$ improvement diminish over the course of several hours. We attribute this to interfacial solidification, which will reduce the effective dielectric constant.

[35] N. Zhao, Z.-Y. Wang, and R.-B. Liu, Phys. Rev. Lett. **106**, 217205 (2011).

[36] P. Huang, X. Kong, N. Zhao, F. Shi, P. Wang, X. Rong, R.-B. Liu, and J. Du, Nat. Comm. **2**, 570 (2011).

[37] K. Fang, V. M. Acosta, C. Santori, Z. Huang, K. M. Itoh, H. Watanabe, S. Shikata, and R. G. Beausoleil, Phys. Rev. Lett. **110**, 130802 (2013).

[38] H. J. Mamin, M. H. Sherwood, M. Kim, C. T. Rettner, K. Ohno, D. D. Awschalom, and D. Rugar, Phys. Rev. Lett. **113**, 030803 (2014).

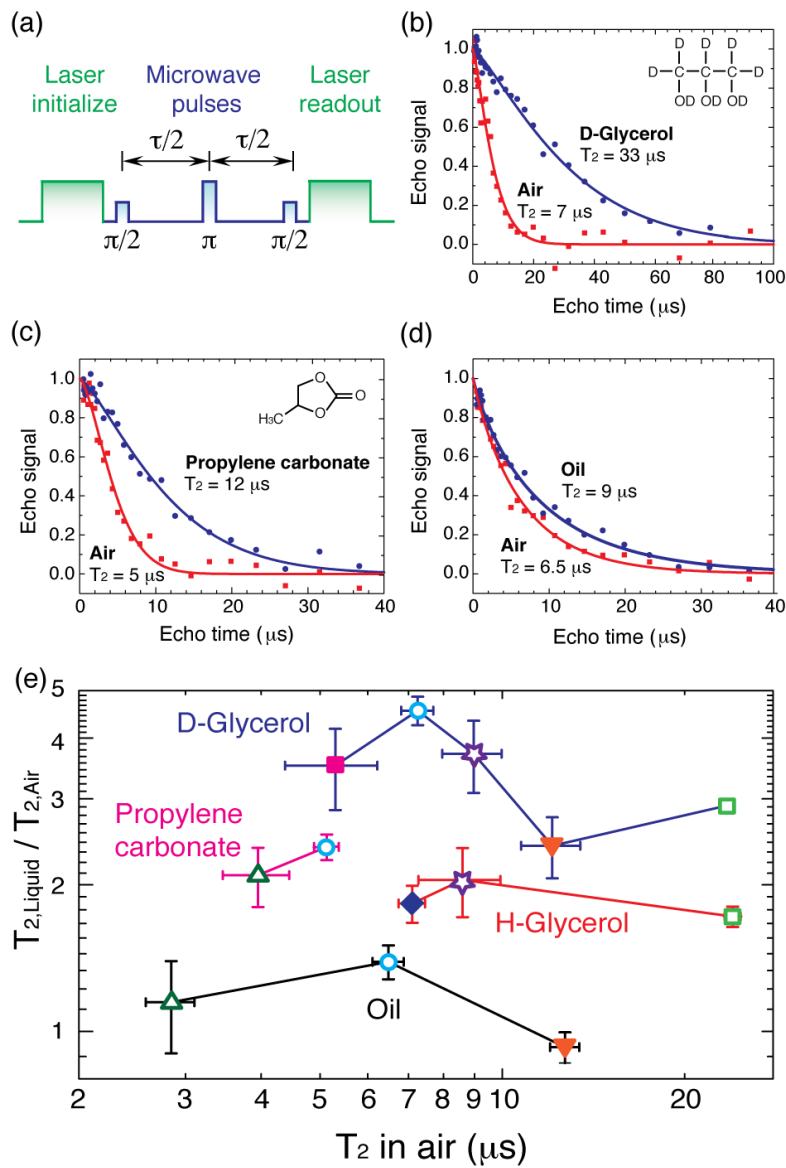

Figure 1 – Effect of various liquids on Hahn echo $T_2$ times. (a) Pulse pattern for optically detected spin echo. (b)-(d) Normalized echo amplitudes obtained in air and with three different liquids covering the diamond surface. The same NV center was used for these three examples. Solid lines are fits to stretched exponentials. Bias field $B_z$ = 39 mT. (e) Summary of $T_2$ ratios. The three liquids with high dielectric constant show a substantial increase in coherence time, with $T_2$ ratios between 1.7 and 4.6. Seven NV centers were tested, with each having a distinct symbol in the plot. Error bars are based on the standard error found from fitting the echo decays.

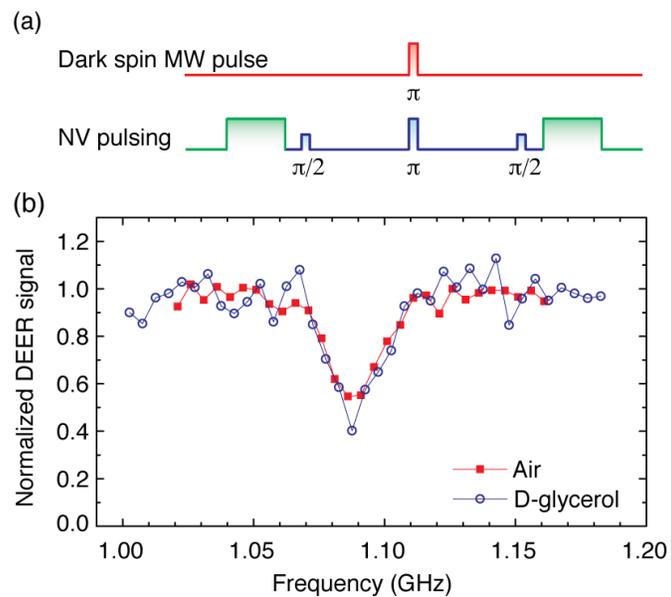

Figure 2 – Double electron-electron spin resonance (DEER) measurements. (a) Pulse pattern for the experiment. The frequency of the dark spin microwave (MW) pulse is scanned and causes spin inversions when the frequency matches the resonance frequency of the dark spins. The spin inversions are detected by their effect on the NV spin echo. (b) DEER measurements for a diamond sample in air and when covered with glycerol. The effect of the dark spin inversions is clearly seen in the dip at 1.09 GHz. Addition of glycerol results in no substantial change in the dark spin signal, indicating that glycerol does not significantly affect the dark spin density. Echo evolution time was 5 µs.

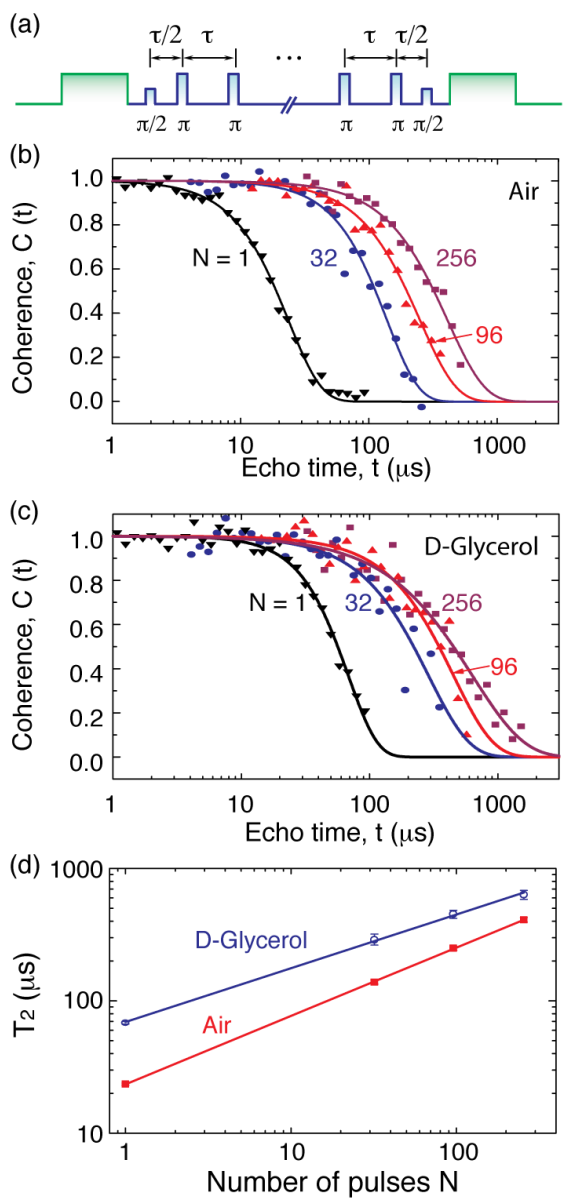

Figure 3 – Results of multipulse spin echo measurements. (a) Pulse sequence for the measurements. The π pulse phases were in an XY8-N pattern, where $N$ is the number of π pulses. (b) Points are coherence data (normalized spin echo amplitudes) taken in air as a function of echo time $t = N\tau$ with a bias field $B_z = 21 \text{ mT}$. Solid curves are fits to stretched exponentials. (c) Same as (b) but measured with deuterated glycerol covering the diamond. (d) $T_2$ as a function of number of π pulses. $T_2$ is proportional to $N^{0.52}$ in air, and $N^{0.41}$ in deuterated glycerol.

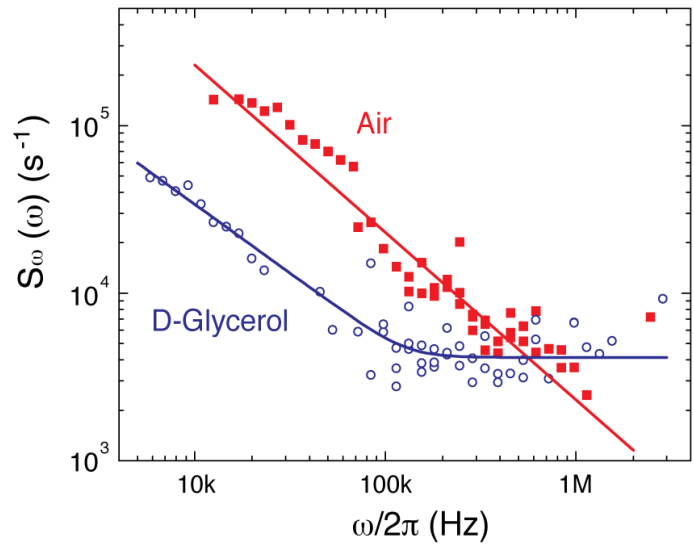

Figure 4 – Power spectral density of precession frequency noise as determined from spectral decomposition of multipulse coherence data. In air, the spectral density falls roughly as $1/\omega$, indicated by solid red line. In deuterated glycerol, the response is approximately $1/\omega^{0.8}$ for frequencies below 100 kHz and levels off for higher frequencies. The solid blue line is a guide for the eye.

# Decoherence of near-surface nitrogen-vacancy centers due to electric field noise

## Supplemental Material

M. Kim, H. J. Mamin, M. H. Sherwood, K. Ohno, D. D. Awschalom and D. Rugar

### 1. Apparatus and detection protocol

This study was performed at room temperature using a custom-built confocal microscope with fluorescence detection, similar to that described previously [1,2], but modified to allow application of liquids to the diamond surface. As shown in Fig. S1, the diamond substrate was placed on top of a 100 μm thick quartz coverglass. A small Teflon-encapsulated O-ring was used to contain the liquid. Laser illumination ($\lambda = 532$ nm) was from the bottom side of the liquid cell with the light focused through both the coverglass and the 160 μm thick diamond substrate. Acceptable compensation for spherical aberration induced by the coverglass and diamond substrate could be obtained using a microscope objective designed for air-incident coverglass correction (Olympus UPLSAPO40X2, NA=0.95).

The spin state of the NV center was initialized and read out using 3 μs laser pulses. During readout, the broadband red fluorescence was detected using appropriate wavelength filtering and single photon counting electronics. A gated photon counter accumulated fluorescence photon counts during the read pulse in two detection windows, which we denote as $s$ (signal) and $r$ (reference). The $s$ count is the number of photons collected during the first 350 ns of the read pulse, and thus measures the spin-dependent fluorescence of the NV center. The $r$ count measures the photon counts later in the read pulse after the NV has been reset back to the $m_s = 0$ state. This enables the signal to be normalized to account for systematic changes in laser power, focus, etc. The pulse sequence was repeated at least 250,000 times in order to collect sufficient photons for acceptable statistics.

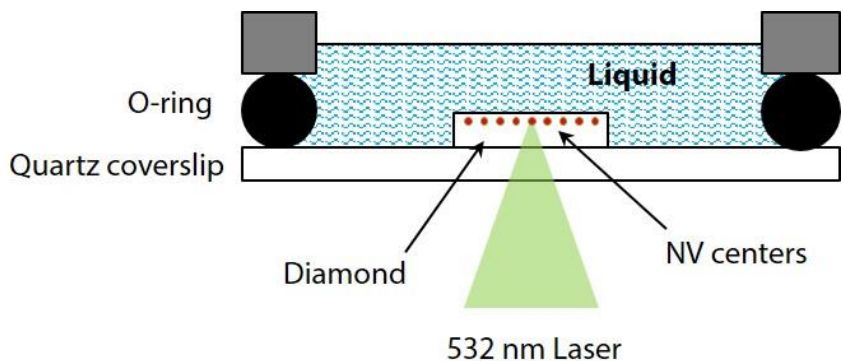

Figure S1 – Diagram of the liquid cell.

The spin echo measurements were always made using two opposite phases for the initial π/2 microwave pulse, which we denote as (π/2)$_{+y}$ and (π/2)$_{-y}$. If there were no decoherence or other interactions, the (π/2)$_{+y}$ initial pulse would leave the NV center in the dark state ($m_s = -1$) at the end of the multipulse sequence, while the (π/2)$_{-y}$ initial pulse would leave the NV in the bright state ($m_s = 0$). We calculate the spin echo contrast from the measured photon counts according to

$$s(\tau) = \frac{s^-(\tau) - s^+(\tau)}{\frac{1}{2}\left[r^-(\tau) + r^+(\tau)\right]} ,\qquad (S1)$$

where $s$ and $r$ indicate the total counts accumulated during the signal and reference gate times. The + and - superscripts indicate photon counts obtained for spin echoes having initial pulses of (π/2)$_{+y}$ and (π/2)$_{-y}$, respectively. Decoherence reduces the spin echo signal contrast as the spin echo evolution time increases. To extract the $T_2$ values, the echo decay curves were fit to the stretched exponential function

$$s(\tau) = s_0 \exp\left[-(\tau/T_2)^n\right].$$

## 2. Sample preparation

Most of the measurements were performed using a commercial (Element Six) electronic-grade [100] oriented diamond substrate that was capped with a 50 nm thick layer of isotropically pure carbon-12 diamond. The capping layer was grown by plasma-enhanced chemical vapor deposition. Near-surface NV centers were created within the capping layer by $^{15}$N ion implantation at 2.5 keV, followed by annealing for six hours at 850°C in vacuum (1×10$^{-9}$ Torr). To remove graphitic contamination and to oxygen terminate the surface, the diamond was cleaned in a 200°C three-acid mixture (equal parts nitric, sulfuric and perchloric acids) for four hours and then heated to 425°C in pure oxygen for two hours [3]. This process results in NV centers that are located roughly 5 nm below the diamond surface.

To verify that the glycerol-induced $T_2$ enhancement was not dependent on the isotopically pure capping layer, measurements were also made using a diamond sample that was prepared without the capping layer (i.e., in diamond with natural abundance of $^{13}$C).

Spin echo measurements were made in air and then after a liquid was applied. To make further measurements, the diamond substrate was then re-cleaned, either by repeating the acid cleaning and oxygen bake procedure, or by washing the sample in boiling water and rinsing with isopropanol.

Table S1 shows the cleaning steps used for each specific measurement that was presented in Fig. 1(e) of the main text. Each NV center has a designated label (e.g., D07, E95, etc.). We reproduce Fig. 1(e) here as Fig. S2, with each data point labeled by the specific NV centers. Some NV centers were used for multiple measurements of $T_2$ with different liquids. As can be seen in Fig. S2 and the fourth column of Table S1, the $T_2$ measured in air reverts back to near its original value after the liquid is removed and the sample re-cleaned.

It is important to take data as soon as possible after application of the glycerol and propylene carbonate. The $T_2$ improvement seen with these liquids diminishes over the course of several hours. We attribute this effect to the formation of a layer of reduced mobility [4,5], essentially a thin semi-solid layer at the diamond surface, which reduces the effective dielectric constant.

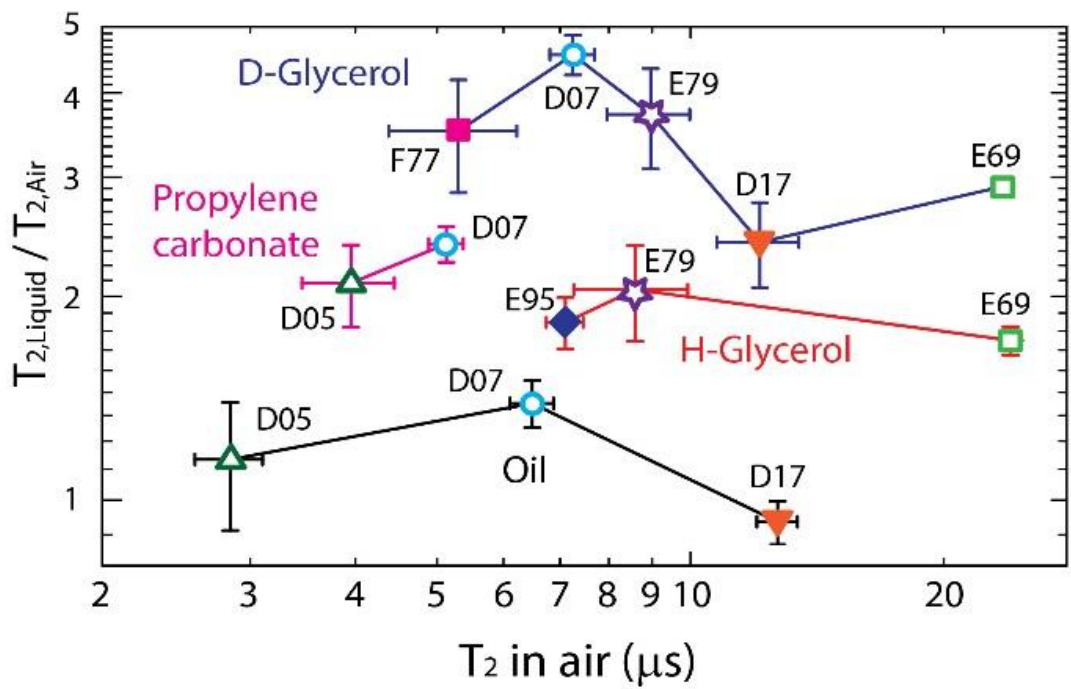

Figure S2 – Effect of various liquids on Hahn echo $T_2$ times. This plot is the same as Fig. 1(e) in the main text except with NV identification labels added.

Table S1 – Process steps for $T_2$ Hahn echo measurements

| NV # | Process step | Surface preparation | $T_2$ (μs) in air | $T_2$ (μs) in liquid | $T_2$ error from fit (μs) | $T_2$ ratio* | Field (mT) |
|---|---|---|---|---|---|---|---|
| D07 | 1 | Acid clean, oxygen bake | 7.3 | | 0.4 | | |
| | 2 | D-Glycerol | | 33 | 1 | 4.6 | 39 |
| | 3 | Boiling water, isopropanol | 5.1 | | 0.24 | | |
| | 4 | Propylene carbonate | | 12.3 | 0.5 | 2.4 | 39 |
| | 5 | Boiling water, isopropanol | 6.5 | | 0.4 | | |
| | 6 | Oil | | 9.0 | 0.5 | 1.4 | 39 |
| D17 | 1 | Acid clean, oxygen bake | 12.1 | | 1.4 | | |
| | 2 | D-Glycerol | | 29.1 | 2.6 | 2.4 | 39 |
| | 3 | Boiling water, isopropanol | 12.7 | | 0.7 | | |
| | 4 | Oil | | 11.8 | 0.5 | 0.93 | 39 |
| D05 | 1 | Acid clean, oxygen bake, boiling water, isopropanol | 4.0 | | 0.5 | | |
| | 2 | Propylene carbonate | | 8.3 | 0.5 | 2.1 | 39 |
| | 3 | Boiling water, isopropanol | 2.8 | | 0.3 | | |
| | 4 | Oil | | 3.3 | 0.6 | 1.1 | 39 |
| E69 | 1 | Acid clean, oxygen bake | 23.5 | | 0.5 | | |
| | 2 | D-Glycerol | | 68.3 | 1.5 | 2.9 | 21 |
| | 3 | Acid clean, oxygen anneal | 24 | | 0.8 | | |
| | 4 | H-Glycerol | | 41 | 1.5 | 1.7 | 21 |
| E79 | 1 | Acid clean, oxygen bake | 9.0 | | 1.0 | | |
| | 2 | D-Glycerol | | 33.3 | 4.2 | 3.7 | 21 |
| | 3 | Acid clean, oxygen bake | 8.6 | | 1.3 | | |
| | 4 | H-Glycerol | | 17.6 | 0.8 | 2.0 | 21 |
| E95 | 1 | Acid clean, oxygen bake | 7.1 | | 0.4 | | |
| | 2 | H-Glycerol | | 13.0 | 0.9 | 1.8 | 21 |
| F77 | 1 | Acid clean, oxygen bake | 5.3 | | 0.9 | | |
| | 2 | D-Glycerol | | 18.6 | 1.4 | 3.5 | 35 |

*$T_2$ ratio is defined as $T_{2,Liquid}/T_{2,air}$, where $T_{2,air}$ is the value measured in the previous process step.

Note all NV centers were formed in a 50 nm thick isotopically pure carbon-12 epilayer, except for F77, which was formed in a commercial electronic-grade substrate with natural abundance of carbon-13. Prior to the above surface preparation steps, the NVs were created with a 2.5 keV $^{15}$N implant followed by vacuum annealing at 850°C.

## 3. Supporting evidence from double quantum measurements

In the main text, we observe the effect of dielectric liquids and conclude that electric field noise contributes significantly to the decoherence of near-surface NV centers. To provide supporting evidence, we consider here a comparison of decoherence between the conventional "single-quantum" (SQ) spin echo and the "double-quantum" (DQ) spin echo, which uses the superposition between the $m_s = +1$ and -1 sublevels [6–9].

For an applied magnetic field along the $z$ axis and to first order in electric field, the energy levels for the $m_s = +1$ and -1 sublevels relative to the $m_s = 0$ sublevel are given by

$$\omega_\pm / 2\pi \approx D + (d_\parallel / h) E_z \pm (\gamma / 2\pi) B_z \tag{S2}$$

where $D = 2.87$ GHz is the zero field splitting. As a consequence of (S2), the coherence of the conventional SQ spin echo (e.g., superposition between $m_s = 0$ and -1 sublevels) is sensitive to fluctuations in any of the parameters: $D$, $B_z$, and $E_z$. In contrast, the DQ spin echo (superposition between $m_s = +1$ and -1 sublevels) will have increased sensitivity to fluctuations in $B_z$, but be insensitive to fluctuations in $D$ and $E_z$ since both $m_s = +1$ and -1 sublevels are affected identically.

To quantify the difference between SQ and DQ echoes, we start by deriving expressions for the echo responses. Starting with the SQ echo, the evolution of the spin state can be described by the product of five rotation operators representing the pulse sequence $\pi/2 - \tau/2 - \pi - \tau/2 - \pi/2$:

$$U_{SQ} = U_x^{SQ}(\pi/2) U(\tau, \tau/2) U_x^{SQ}(\pi) U(\tau/2, 0) U_x^{SQ}(\pi/2). \tag{S3}$$

In matrix form, assuming we are using the 0 and -1 magnetic sublevels, the ideal SQ pulse operators are:

$$U_x^{SQ}(\pi/2) = \begin{pmatrix} 1 & 0 & 0 \\ 0 & \frac{1}{\sqrt{2}} & \frac{-i}{\sqrt{2}} \\ 0 & \frac{-i}{\sqrt{2}} & \frac{1}{\sqrt{2}} \end{pmatrix} \tag{S4}$$

$$U_x^{SQ}(\pi) = \begin{pmatrix} 1 & 0 & 0 \\ 0 & 0 & -i \\ 0 & -i & 0 \end{pmatrix}. \tag{S5}$$

The evolution operator in the first half of the echo from time $t = 0$ to $\tau/2$ is

$$U(\tau/2,0) = \begin{pmatrix} \exp[-i(\phi_E + \phi_B)] & 0 & 0 \\ 0 & 1 & 0 \\ 0 & 0 & \exp[-i(\phi_E - \phi_B)] \end{pmatrix}, \quad (S6)$$

where $\phi_E = 2\pi \int_0^{\tau/2} [D + (d_\parallel/h) E_z(t)] dt$ and $\phi_B = \gamma \int_0^{\tau/2} B_z(t) dt$. Likewise, for the second half of the echo

$$U(\tau,\tau/2) = \begin{pmatrix} \exp[-i(\phi'_E + \phi'_B)] & 0 & 0 \\ 0 & 1 & 0 \\ 0 & 0 & \exp[-i(\phi'_E - \phi'_B)] \end{pmatrix}, \quad (S7)$$

where $\phi'_E = 2\pi \int_{\tau/2}^{\tau} [D + (d_\parallel/h) E_z(t)] dt$ and $\phi'_B = \gamma \int_{\tau/2}^{\tau} B_z(t) dt$.

Assuming that the echo sequence starts in the $m_s = 0$ state, the probability for ending back in that state is given by

$$P_0^{SQ}(\tau) = |\langle 0|U_{SQ}|0\rangle|^2 = \cos^2\left(\frac{\Delta\phi_E - \Delta\phi_B}{2}\right)$$
$$= \frac{1}{2} + \frac{1}{2}\cos(\Delta\phi_E - \Delta\phi_B) \quad (S8)$$

where $\Delta\phi_E = \phi_E - \phi'_E$ and $\Delta\phi_B = \phi_B - \phi'_B$. This result confirms that the single-quantum spin echo is sensitive to fluctuations in both the electric and magnetic fields.

If $\Delta\phi_E$ and $\Delta\phi_B$ are uncorrelated Gaussian random variables, then one can show that the average value of $P_0^{SQ}$ is

$$\langle P_0^{SQ} \rangle = \frac{1}{2} + \frac{1}{2}\exp\left[-\frac{\left(\langle(\Delta\phi_E)^2\rangle + \langle(\Delta\phi_B)^2\rangle\right)}{2}\right]. \quad (S9)$$

The spin echo signal normalized between the values of 0 and 1 is given by

$$s_{SQ}(\tau) = 2\langle P_0^{SQ}\rangle - 1$$
$$= \exp\left[-\frac{\left(\langle(\Delta\phi_E)^2\rangle + \langle(\Delta\phi_B)^2\rangle\right)}{2}\right] \tag{S10}$$

We now repeat the analysis for the case of the double-quantum echo. The double-quantum echo is the result of the five evolution operators

$$U_{DQ} = U_x^{DQ}(\pi/2)U(\tau,\tau/2)U_x^{DQ}(\pi)U(\tau/2,0)U_x^{DQ}(\pi/2), \tag{S11}$$

where the ideal DQ pulse operators are [9]

$$U_x^{DQ}(\pi/2) = \begin{pmatrix} \frac{1}{2} & \frac{-i}{\sqrt{2}} & -\frac{1}{2} \\ \frac{-i}{\sqrt{2}} & 0 & \frac{-i}{\sqrt{2}} \\ -\frac{1}{2} & \frac{-i}{\sqrt{2}} & \frac{1}{2} \end{pmatrix} \tag{S12}$$

and

$$U_x^{DQ}(\pi) = \begin{pmatrix} 0 & 0 & -1 \\ 0 & -1 & 0 \\ -1 & 0 & 0 \end{pmatrix}. \tag{S13}$$

The result of the DQ echo sequence can now be evaluated:

$$P_0^{DQ}(\tau) = |\langle 0|U_{DQ}|0\rangle|^2 = \cos^2(\Delta\phi_B)$$
$$= \frac{1}{2} + \frac{1}{2}\cos(2\Delta\phi_B) \tag{S14}$$

Treating $\Delta\phi_B$ as a Gaussian random variable, we get

$$\langle P_0^{DQ}\rangle = \frac{1}{2} + \frac{1}{2}\exp\left[-2\langle(\Delta\phi_B)^2\rangle\right] \tag{S15}$$

and find the corresponding normalized spin echo signal to be

$$s_{DQ}(\tau) = 2\langle P_0^{DQ}\rangle - 1$$
$$= \exp\left[-2\langle(\Delta\phi_B)^2\rangle\right] \quad (S16)$$

By comparison to the SQ case in (S10), we see that the DQ spin echo is not affected by fluctuations in $E_z$, but is four times more sensitive to the mean square fluctuations of $B_z$. This increase in sensitivity to magnetic fluctuations is in agreement with the finding of Zhao et al. [6].

To directly compare SQ and DQ echoes, we consider the following ratio

$$r(\tau) = \frac{\log s_{SQ}(\tau)}{\log s_{DQ}(\tau)}$$
$$= \frac{\langle(\Delta\phi_E)^2\rangle + \langle(\Delta\phi_B)^2\rangle}{4\langle(\Delta\phi_B)^2\rangle} \quad (S17)$$

If the only source of decoherence is magnetic field noise, then $r = 0.25$. If electric field fluctuations are significant, then we expect to find $r > 0.25$. If $r$ is known from measurements, then the contribution of the electric field relative to the magnetic noise is given by

$$\frac{\langle(\Delta\phi_E)^2\rangle}{\langle(\Delta\phi_B)^2\rangle} = 4r - 1. \quad (S18)$$

We now turn our attention to experimental results that compare single- and double-quantum spin echoes. The double-quantum methodology we use has been described previously [9], and is based on dual frequency microwave pulses which simultaneously address both the 0 to -1 and 0 to +1 transitions. In that previous paper [9], we presented experimental SQ and DQ spin echo decay data using a rather deep NV center ( ~12 nm deep). We include an analysis of that data here, as well as new measurements using two of the shallower NVs available in the present sample (~ 5 nm deep). In all cases, we found $r > 0.25$, suggesting significant contribution of electric field noise.

Figure S3 shows SQ and DQ spin echo data for the three NV centers and the corresponding results for the ratio $r(\tau)$. For the ~12 nm deep NV shown in Figs. S3(a) and (b), we find $r \approx 0.33$. From (S18), this indicates a relatively modest contribution from electric field noise: $\langle(\Delta\phi_E)^2\rangle / \langle(\Delta\phi_B)^2\rangle \approx 0.3$.

A considerably stronger effect was found for the two shallower NV centers. For example, in Fig. 3(f), the $r$ value is about 0.45 for $\tau$ in the vicinity of 5 μs. From this value of $r$, we find $\langle(\Delta\phi_E)^2\rangle / \langle(\Delta\phi_B)^2\rangle \approx 0.8$, indicating that, in this case, electric field noise is indeed a significant contributor to the decoherence.

Finally, we note that this analysis has ignored the effect of fluctuations in $E_\perp^2$ (see equation (3) in main text). Since fluctuations in $E_\perp^2$ act on the spin state energy levels in a manner similar to fluctuations in $B_z$, they would not be obvious in a comparison of DQ and SQ spin echoes. Consequently, if fluctuations in $E_\perp^2$ are significant, the above analysis would underestimate the effect of electric field noise.

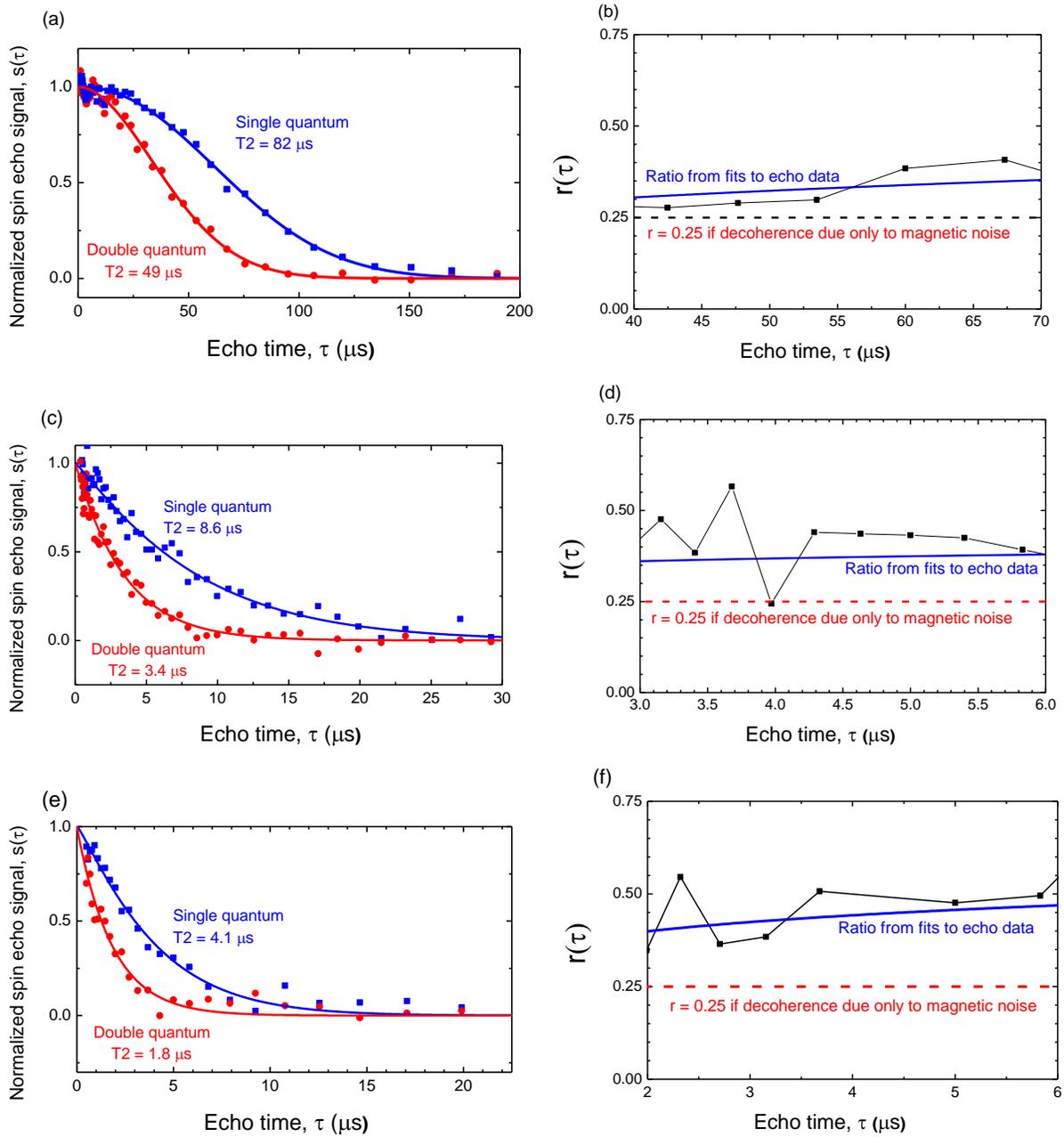

Figure S3 – Single- and double-quantum spin echo comparisons for three different NV centers in air (no applied liquids). (a), (c), (e) Points are measured spin echo signals. Solid curves are fits to stretched exponential decays. Data points in (a) are from Ref. [9] using a ~12 nm deep NV center. Data in (c) and (e) are from ~5 nm deep NVs. (b), (d), (f) Ratio of signal logarithms, $r(\tau)$, from equation (S17). Points are ratios calculated using the measured spin echo signal data. Solid blue curves are calculated from fitted spin echo decay curves. Dashed red line shows the $r = 0.25$ value expected if decoherence is due solely to magnetic field noise. The fact that $r > 0.25$ suggests that electric field noise contributes to the decoherence for each NV center, with greater effect in the shallower NVs.

# References

[1]  D. Rugar, H. J. Mamin, M. H. Sherwood, M. Kim, C. T. Rettner, K. Ohno, and D. D. Awschalom, Nat. Nanotech. **10**, 120 (2015).

[2]  H. J. Mamin, M. Kim, M. H. Sherwood, C. T. Rettner, K. Ohno, D. D. Awschalom, and D. Rugar, Science **339**, 557 (2013).

[3]  M. Kim, H. J. Mamin, M. H. Sherwood, C. T. Rettner, J. Frommer, and D. Rugar, Appl. Phys. Lett. **105**, 042406 (2014).

[4]  L. Xu and M. Salmeron, J. Phys. Chem. B **102**, 7210 (1998).

[5]  S. Capponi, S. Napolitano, N. R. Behrnd, G. Couderc, J. Hulliger, and M. Wübbenhorst, J. Phys. Chem. C **114**, 16696 (2010).

[6]  N. Zhao, Z.-Y. Wang, and R.-B. Liu, Phys. Rev. Lett. **106**, 217205 (2011).

[7]  P. Huang, X. Kong, N. Zhao, F. Shi, P. Wang, X. Rong, R.-B. Liu, and J. Du, Nat. Comm. **2**, 570 (2011).

[8]  K. Fang, V. M. Acosta, C. Santori, Z. Huang, K. M. Itoh, H. Watanabe, S. Shikata, and R. G. Beausoleil, Phys. Rev. Lett. **110**, 130802 (2013).

[9]  H. J. Mamin, M. H. Sherwood, M. Kim, C. T. Rettner, K. Ohno, D. D. Awschalom, and D. Rugar, Phys. Rev. Lett. **113**, 030803 (2014).